\documentclass[]{spie}

\usepackage{amsmath,amsfonts,amssymb}
\usepackage{graphicx}
\usepackage[colorlinks=true, allcolors=blue]{hyperref}

\usepackage[per-mode=fraction, quotient-mode=fraction, separate-uncertainty=true]{siunitx}
\DeclareSIUnit\muev{\micro\electronvolt} 

\title{Interfacing superconducting nanowire single photon detectors with cryogenic opto-electronics for quantum photonic applications}

\author[a,b]{Niklas Lamberty}
\author[a,b]{Frederik Thiele}
\author[b]{Thomas Hummel}
\author[a,b]{Tim J. Bartley}
\affil[a]{Department of Physics, Paderborn University, Warburger Str. 100, 33098 Paderborn, Germany}
\affil[b]{Institute for Photonic Quantum Systems (PhoQS), Paderborn University, Warburger Str. 100, 33098 Paderborn, Germany}

\authorinfo{(Send correspondence to N.L.)\\N.L.: E-mail: niklas.lamberty@uni-paderborn.de}

\begin{document} 
\maketitle

\begin{abstract}
Interfacing single-photon detectors with active photonic components is a cornerstone photonic quantum technology. We describe how the output signal of commercial superconducting nanowire single-photon detectors can be used \emph{in situ} to drive photonic components such as lasers and electro-optic modulators, co-located in the cryostat. This is enabled by developing custom circuitry using cryogenic-compatible discrete components in the SiGe-BiCMOS platform. We have demonstrated this with a number of experiments, in particular optical readout of an SNSPD and low-latency feed-forward modulation based on single-photon measurement events, all at or below \qty{4}{K}. This manuscript is an abridged version of the Master thesis of the primary author N. Lamberty.
\end{abstract}


\section{INTRODUCTION}
\label{sec:intro}
\begin{figure}[ht]
   \begin{center}
   \begin{tabular}{c}
   \includegraphics[width=\textwidth]{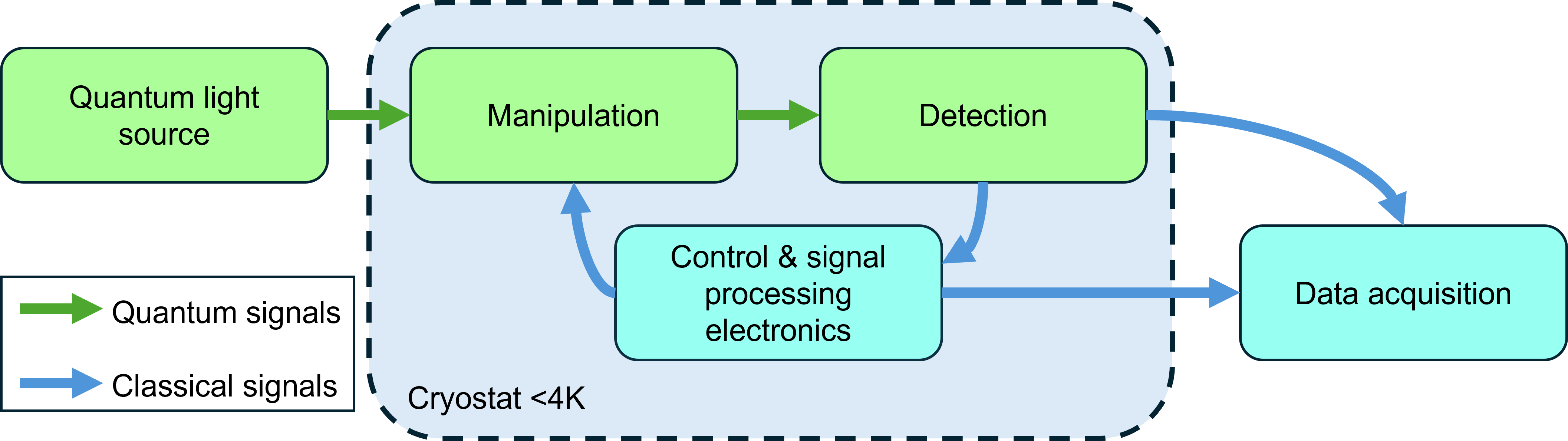}
   \end{tabular}
   \end{center}
   \caption[example] 
   { \label{fig:schematic} 
Schematic of a general photonic quantum system, comprising a quantum light source, manipulation and detection. Detection and manipulation may be linked by classical signals via control and signal processing electronics. In our scheme, manipulation, detection and control are all co-located at \SI{<4}{K}, the operating temperature of the superconducting detectors.}
   \end{figure}

Photonic quantum systems are based on interfacing quantum light sources, manipulation and measurement at the single-photon level\cite{Slussarenko2019}. Key to preserving nonclassical properties of quantum light, and maintaining any quantum advantage in information processing tasks, is to use high-performance components and minimize losses at interfaces between them.
Many high-performance quantum optical technologies require cryogenic temperatures, such as single-photon emitters and superconducting nanowire single-photon detectors (SNSPDs) \cite{Migdall2013}. Therefore, to minimize processing time, control and signal processing electronics must be combined with photonic quantum state manipulation alongside the superconducting detectors in a cryostat. This concept is shown schematically in Fig.~\ref{fig:schematic}. Recently, we have shown how nonlinear and electro-optic components based on titanium in-diffused waveguides lithium niobate can be operated under cryogenic conditions\cite{Thiele2020,Bartnick2021,Lange2022,Thiele2022,Thiele2023,Lange2023,Thiele2024b}. Nevertheless, interfacing combinations of components requires additional engineering steps.

Combining SNSPDs with electro-optic modulators poses unique challenges to bridge the voltage gap between the detection signal and modulation voltage of opto-electronic components. 
While SNSPDs generate sub-\unit{mV} analog signals, our modulators typically require digital signals in the range of \qty{3}{V}~\cite{Thiele2024b}. Creating such an interconnect purely with passive electrical components has been shown to suffer from a steep reduction in speed, extinction ratio and available logical operations~\cite{Thiele2023}. We attempt to bridge this gap using conventional semiconductor electronics instead. This allows us to use the wide range of commercial semiconductor components.

In this paper, we provide details on the cryogenic electronics used to connect the output of an SNSPD to other photonic components. All circuitry is based on combining discrete components on printed circuit boards (PCBs) for efficient prototyping and cryogenic testing. In Section 2 we briefly discuss the effects on material properties when operating semiconductor devices at \qty{4}{K}. In Section 3 we describe the analog electronics used for amplification of the initial SNSPD signal, and show that this can be used to read out the photon-number from an SNSPD signal as well as drive a cryogenic laser diode~\cite{Thiele2024}. In Section 4 we describe cryogenic digital electronics, which we use to read out a multipixel device based on logical signal processing, and drive an integrated electro-optic modulator. This is the basis of our quantum photonic feed-forward experiment~\cite{Thiele2024b}. We conclude in Section 5 with a brief outlook.

\section{Semiconductor devices at 4K}
In order to bridge the gap between the SNSPDs and electro-optic elements intermediate electronic processing is required. We focus on conventional semiconductors, but note that there have been promising demonstrations of superconducting logic based on rapid single flux quantum~\cite{Miki2021} and nano cryotron~\cite{Huang2024} circuits.
\subsection{Silicon germanium heterojunction bipolar transistors (SiGe HBTs)}
In the context of cryogenic quantum photonics and opto-electronics, Silicon-Germanium Hetero-Junction-Bipolar Transistors (SiGe HBTs) and Silicon MOSFETs are promising devices, since they can be co-integrated in
commercial foundries and show good performance at \qty{4}{K}.  SiGe-HBTs are a type of bipolar transistors, which feature a SiGe alloy in the base. This allows for higher doping densities, due to the intrinsic band offset between silicon and
silicon germanium supporting the bandstructure of the n-p-n transistor. The graded germanium alloy in the base also creates a graded conduction band profile, which in turn creates an electric field in the base. This field helps to flush carriers out of the base region, which increases the speed of these devices. The more efficient drift transport due to the graded band also decreases the base current and thereby increases the current gain~\cite{Cressler1998}. This type of technology has already been used to perform fast electrical control of optical elements at \qty{4}{K}~\cite{Widhalm2018} and is thus a proven tool in this field.

SiGe-HBTs have been shown to have good performance at \qty{4}{K}~\cite{Rucker2017} with an increased transconductance and higher current gain. Operating these transistors at \qty{4}{K} give rise to effects like tunneling~\cite{Ying2018} and the kink effects~\cite{Liang2006}, which cannot be seen at room temperature and make the transfer characteristic of the transistors less linear. Another change is a higher turn-on voltage caused by band gap shifting~\cite{Rucker2017}, due to thermal contraction of the crystal lattice. The doping densities used in these devices are generally high enough to operate in the degenerate regime, which starts at \qty{\sim1d20}{cm^{-3}}~\cite{Chapman1963}, where the energy level of the dopant blends with the conduction band~\cite{Marshak1984}. This prevents freeze-out of carriers at low temperatures~\cite{Chapman1963} as carriers no longer need to be thermally activated.

Compared to room temperature operation, the transport of carriers in the base region of SiGe-HBTs is fundamentally different. Under ambient conditions, transport is well described by drift-diffusion theory. Since diffusion is a thermally activated process, this theory would thus predict a very small base current under cryogenic conditions due to drift-diffusion transport. In reality, significant deviations from classical theory are observed~\cite{Rucker2017}. This is due to effects like trap assisted tunneling~\cite{Ying2018}, whereby electrons tunnel from the emitter to the collector with intermediate stops at deep trap levels within the base. Such trap levels can be caused by impurities like copper or gold during manufacturing. At higher currents quasi ballistic transport, where electrons pass the base without scattering, becomes a significant contribution~\cite{Ying2018}. Theoretical descriptions therefore need to account for many new factors, which makes accurate simulations of the behavior of these transistors at \qty{4}{K} difficult.

\subsection{Metal-oxide-semiconductor field-effect transistors (MOSFETs)}
Another type of transistor which also works at \qty{4}{K} is the MOSFET. This type of transistor is also made up of a p-n-p (p-mos) or n-p-n (n-mos) junction~\cite{JAIN1987}, as is the case for bipolar transistors. Unlike the bipolar transistor the middle region is not directly contacted but capped with an insulator layer before the deposition of the gate contact.

At 4 K MOSFETs exhibit a higher turn on voltage due to band shifting and higher conducted currents due to increased carrier mobility~\cite{VanDijk2020}. The variance between individual transistors also increases. This makes more delicate analog circuits harder to design~\cite{VanDijk2020}. We therefore use MOSFETs only in roles requiring digital signal processing. The overall transport mechanism of these transistors remains unchanged at \qty{4}{K}. This makes them a good candidate for cryogenic electronics with large scale cryo-CMOS circuits already showing potential as controllers for quantum computers~\cite{VanDijk2020}.

\section{Analog SNSPD Readout Interfaces}
\label{sec:title}
Conventionally, the electrical signal from an SNSPD is transmitted to room-temperature with coaxial cables and then amplified with room temperature electronics. We now move this amplification step inside the cryostat, and show two applications for this cryogenic amplification. One is the readout of photon number information from a single SNSPD. We then show the feasibility of driving electro-optics inside the cryostat by driving a laser diode with the amplifier~\cite{Thiele2024}. This electro-optic transduction transfers the SNSPD click signal from the cryostat to room temperature with a fiber optic link.

\subsection{First Stage Amplifier}\label{sec:fsamp}
\begin{figure}
	\centering
	\includegraphics[width=0.9\linewidth]{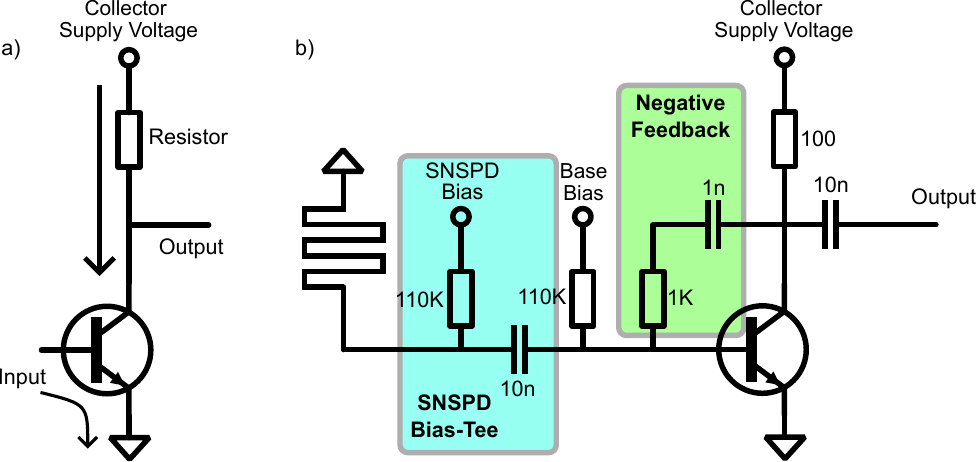}
	\caption{a) A schematic of a simple low noise amplifier made up of a bipolar junction transistor and a resistor. The arrows indicate the direction of current flow. b) The schematic of the first stage amplifier. The SNSPD bias-tee is comprised of a large resistor to convert a voltage in the hundreds of milivolts to a microampere current and a large capacitor for wideband AC-coupling to the amplifier. Negative feedback for the transistor is provided by a resistor capacitor loop.}
	\label{fig:firststageamp}
\end{figure}

For the initial amplification of the SNSPD signals we use a simplified amplifier design based on a single transistor, as shown in Fig.~\ref{fig:firststageamp} a). The simplification enables a low power amplification with low noise. In order to match the input impedance of the transistor base to the coaxial line impedance (usually \qty{50}{\Omega}) additional biasing is needed, which is achieved with a bias tee at the input. Due to the diode-like behavior of the base-emitter junction, changing the bias current over this junction changes the slope of the I-V-curve and thereby the resistance. The collector of the transistor also requires voltage biasing to function. Here, a larger voltage can increase the high frequency performance due to reduced junction capacitances, but this comes at the cost of higher heat dissipation~\cite{Montazeri2016}.

\subsubsection{Circuit design}
One design requirement for the first amplifier is a low heatload due to its direct cable connection from the \qty{4}{K} stage to the \SI{1}{K} stage of the cryostat, which has approximately \SI{500}{\micro W} of cooling power and is thus very sensitive to heat leakage from the amplifier. Another important aspect is a low input reflectivity over a broad frequency range as signal reflections can cause the SNSPD to latch or cause afterpulsing~\cite{Burenkov2013}. Lastly in our case a good reverse isolation, which describes the attenuation of signals propagating from the output to the input of the amplifier, is also required as the readout lines to room-temperature run close to noisy pump systems in the upper cooling stages. The reverse isolation prevents this noise from reaching the SNSPD and causing dark counts.

Circuit design for \SI{4}{K} poses some subtle challenges for amplifier design. One important aspect is to avoid the creation of parasitic LC resonator circuits as the resistivity of copper drops by a factor of 5~\cite{Cahill2016} at cryogenic temperatures. The subsequent increase in Q-factor for these resonators can cause resonances that cannot self-oscillate at room temperature, but do self-oscillate at \SI{4}{K}.
One common way to avoid this is to place small resistors between potential oscillation sources to dampen the oscillations. This is done for all larger capacitors of \SI{100}{nF} or more in all subsequent circuits.

The amplifier itself is designed in a standard LNA configuration with a negative feedback loop to stabilize the amplifier. The full circuit configuration can be seen in Fig.~\ref{fig:firststageamp} b). Since this amplifier is meant to interface directly with the SNSPD, the SNSPD's current source is integrated on the same PCB. The AC-coupling is done with relatively large \SI{10}{nF} NP0 ceramic capacitors to achieve a low reflectivity down to lower frequencies. Decoupling capacitors are omitted in all circuit diagrams but are present on every DC-voltage to reduce noise and stabilize the voltages.

\subsubsection{Characterization}
The amplifier is characterized using a Vector network analyzer (VNA) (PicoVNA-108). This device measures the S-Parameters of the amplifier, which describe the input reflection(S11), forward gain(S21), reverse isolation(S12) and output reflection(S22). The VNA has a measurement range from \SI{300}{kHz} to \SI{8.5}{GHz} with a minimum power of \SI{-20}{dBm}. 
The measurement power of the VNA has to be additionally reduced with an attenuator to get to a power of \SI{-40}{dBm} which is roughly the expected power from the SNSPD. This however makes the measurement somewhat noisy especially for S11 which passes the attenuator twice. The losses from cabling in the cryostat are calibrated out of the data.

\begin{figure}
	\centering
	\includegraphics[width=0.9\linewidth]{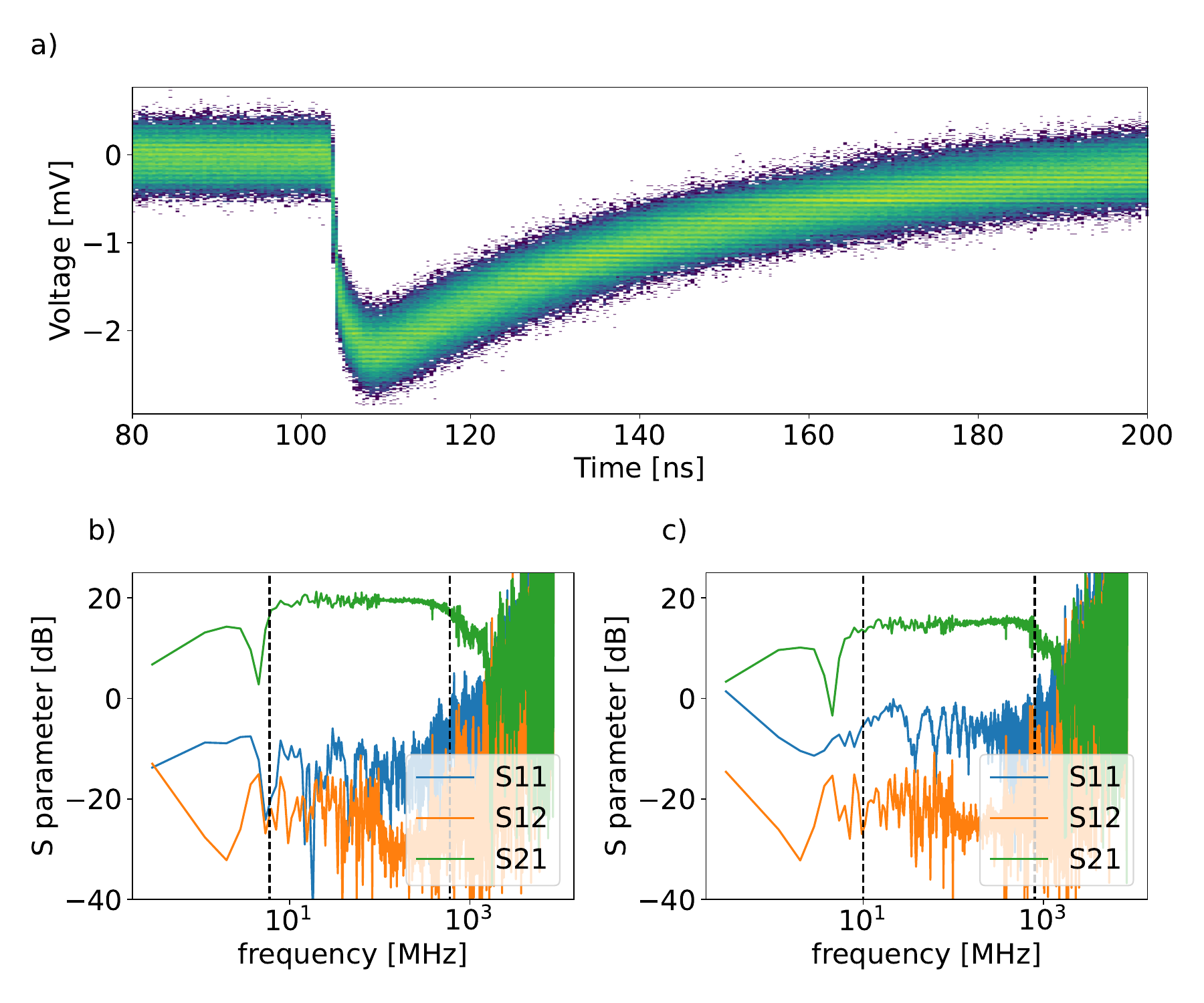}
	\caption{a) A heatmap of traces recorded from a single pixel SNSPD, which is amplified using the first stage amplifier. The primary noise contribution is from the oscilloscope due to the small signal level. b) the S-parameters for the first stage amplifier measured at \SI{4}{K} with a high power setting of \SI{1.4}{mW} of power dissipation. The S-parameters describe the input reflection (S11), the forward gain (S21) and the reverse isolation (S12). The black lines indicate the bandwidth of the amplifier. c) the S-parameters for the first stage amplifier measured at \SI{4}{K} with a low power setting of \SI{0.3}{mW} of power dissipation. }
	\label{fig:amplifiersparamtrace}
\end{figure}

The amplifier can be biased at different base currents and collector voltages to achieve lower power dissipation at the cost of some performance. The measured S-Parameters in Fig.~\ref{fig:amplifiersparamtrace} b) are for a high power setting, dissipating \SI{1.3}{mW}. The amplifier achieves a gain bandwidth from \SI{6}{MHz} to \SI{600}{MHz} and low input reflectivity of about \SI{-10}{dB} for an even greater bandwidth.
For a lower power dissipation of around \SI{300}{\micro W}, as seen in Fig.~\ref{fig:amplifiersparamtrace} the amplifier has a reduced gain of around \SI{15}{dB} and a higher reflectivity. A power dissipation of \SI{300}{\micro W} would nevertheless be compatible with operation directly inside the \SI{1}{K} stage of the cryostat allowing for closer integration.
However, we operate the circuit exclusively at the \SI{4}{K} stage of the cryostat, as this is where all other electronics are located.

\begin{figure}
	\centering
	\includegraphics[width=0.9\linewidth]{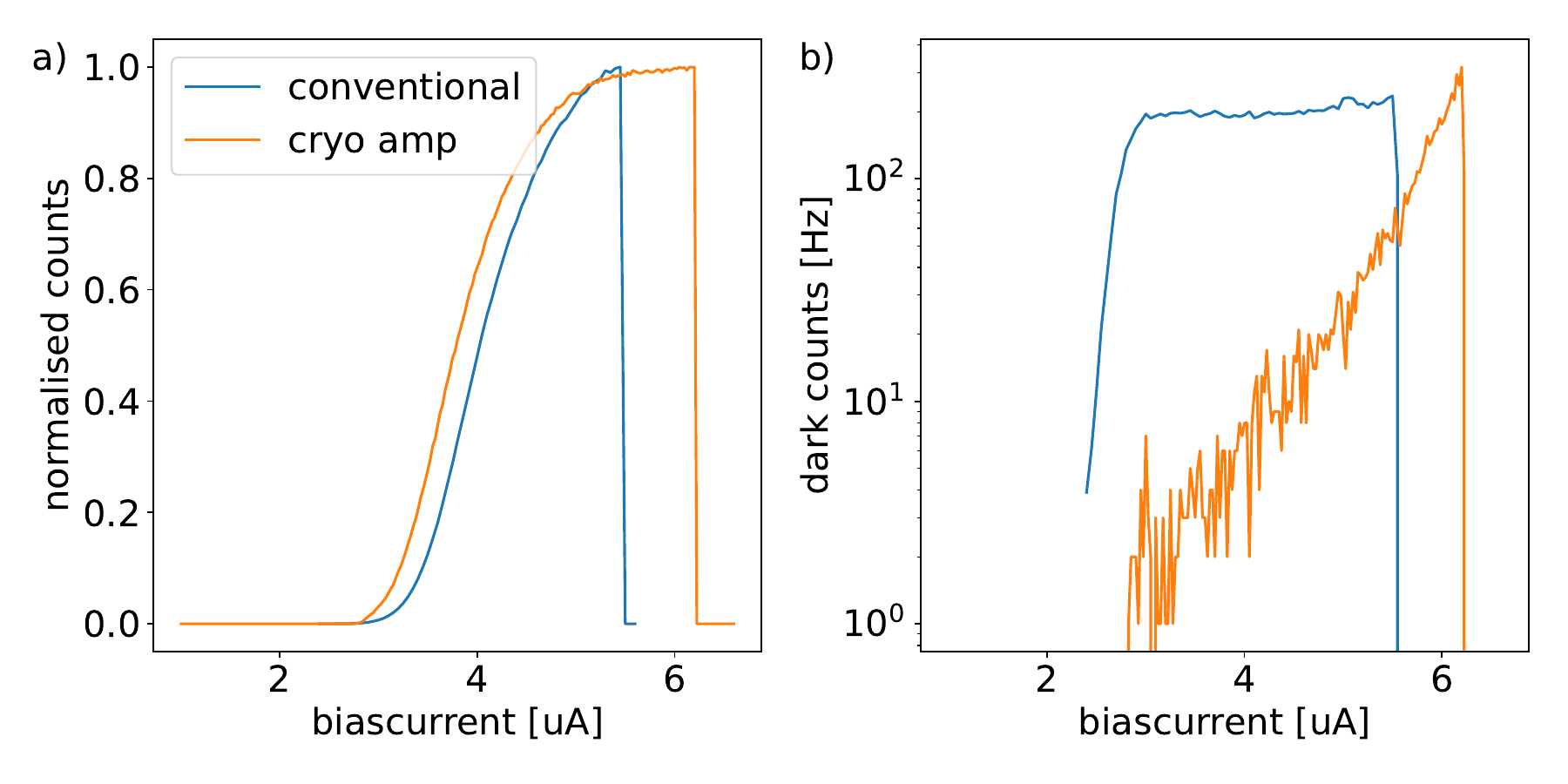}
	\caption{a) The count rate of the SNSPD when illuminated with an attenuated coherent state under different bias currents. The operation with a cryogenic amplifier shows a longer plateau of the count rate and a higher latching current. b) The dark-count rate of the detector when operated with a room temperature (conventional) readout or cryogenic amplification and biasing.}
	\label{fig:stabilityenhancement}
\end{figure}

Another important feature of the amplifier is its effect on the stability of the SNSPD. A scan of the SNSPD bias current while measuring the observed count rate under illumination with a weak coherent state can be seen in Fig.~\ref{fig:stabilityenhancement}. Compared to room temperature amplification, the plateau region before the detector latches is extended and the dark-counts are reduced. This is likely because the SNSPD is more strongly isolated from noise generated by the cold head of the cryostat and rf-noise present in the environment outside.

The power dissipation of \SI{1.3}{mW} is in the range of other cryogenic low noise amplifiers as can be seen in Fig.~\ref{tab:ampcomp}. Some of these amplifiers are however manufactured in a dedicated manufacturing process at a commercial foundry, where the designers have control over transistor dimensions, in contrast to our approach which utilizes off the shelf transistors. Some low power amplifiers also employ spiking of the output using inductors, which is not feasible for lower frequencies and severely limits their bandwidth.

    \begin{table}\centering
\begin{tabular}{|l|c|c|c|c|c|}
	\hline
	 & Power dissipation  & lower BW& upper BW& Gain& discrete \\
	 & [mW] & [MHz] & [MHz] & [dB]& components \\
	\hline
	Liu et al. \cite{Liu2016} & 9.6 & 200 & 4000 & 30 & yes  \\
	\hline
	Montazeri at al. \cite{Montazeri2016} & 0.36 & 1800 & 3600 & 27 & yes  \\
	\hline
	Niu et al. \cite{Niu2022} & 1.8 & 0.2 & 1000 & 26 & no  \\
	\hline
	Li et al. \cite{Li2021} & 3.1 & 0.12 & 1300 & 20.5 & no  \\
	\hline
	this work & 1.3 & 6 & 600 & 20 & yes  \\
	\hline
\end{tabular}\label{tab:ampcomp}
\caption{A comparison of other cryogenic amplifiers to the cryogenic amplifier designed in this work. The amplifiers described by Niu et al.\cite{Niu2022} and Li et al.\cite{Li2021} were explicitly designed to be used with SNSPDs.}
\end{table}

\subsubsection{Future improvements}
One possible improvement would be to DC-couple the amplifier to the SNSPD without the use of a capacitor. This has been demonstrated to increase the maximum count rate of the detector~\cite{Kerman2013}. Previous DC-coupled amplifiers were however built with HEMTs instead of HBTs, which are easier to DC-couple because they do not require a bias current. Another improvement would be the use of a smaller area transistor. This could reduce the heatload of the device, since the required bias current for impedance matching and high gain scales with transistor area. Lastly a high impedance readout near the SNSPD could also increase the signal amplitudes by more closely matching the roughly \qty{1}{k\Omega} source impedance of the SNSPD instead of using a \qty{50}{\Omega} transmission line between the SNSPD and amplifier.

\subsection{Application: intrinsic photon-number resolution}\label{ch:PNR}

One of the primary advantages of using a cryogenic amplifier is the reduced noise at \SI{4}{K}~\cite{Bardin2021}. The low noise and high bandwidth readout which is possible with such an amplifier enables the analysis of smaller features of the SNSPD signal. One such feature is the slope of the rising edge and falling edge of the signal, which has been shown to contain photon number information~\cite{Sauer2023,Schapeler2024}.

\begin{figure}
	\centering
	\includegraphics[width=0.9\linewidth]{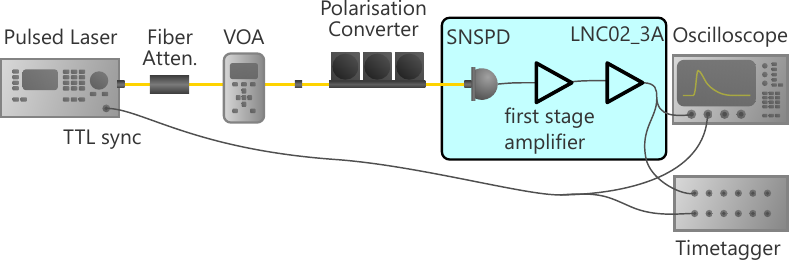}
	\caption{The setup for measuring photon-number resolution with a single-pixel SNSPD. The pulsed laser is attenuated down to single-photon level and send to an SNSPD. The polarization converter is used to optimize the efficiency of the SNSPD. The click signal is amplified by the first stage amplifier and the commercial amplifier. Measurements on the resulting signal are performed either with an oscilloscope or a time-to-digital converter.}
	\label{fig:pnrsetup}
\end{figure}

In order to measure photon-number resolution (PNR), we use the setup shown in Fig.~\ref{fig:pnrsetup}. Since the intrinsic PNR is only visible for short pulses, a pulsed laser diode (PICOPOWER-LD-1550-50-FC) with a pulse duration of \SI{9}{ps} is used to illuminate the SNSPD. The signal level from only the first stage amplifier is not large enough to be visible on a time-to-digital converter. We thus employ an additional commercial cryogenic amplifier (LNF-LNC02\_3A) to further boost the signal level. The resulting detection signal are then analyzed on an oscilloscope (Teledyne T3PS23203) or time to digital converter (Swabian Instruments Time Tagger Ultra).

\begin{figure}
	\centering
	\includegraphics[width=0.8\linewidth]{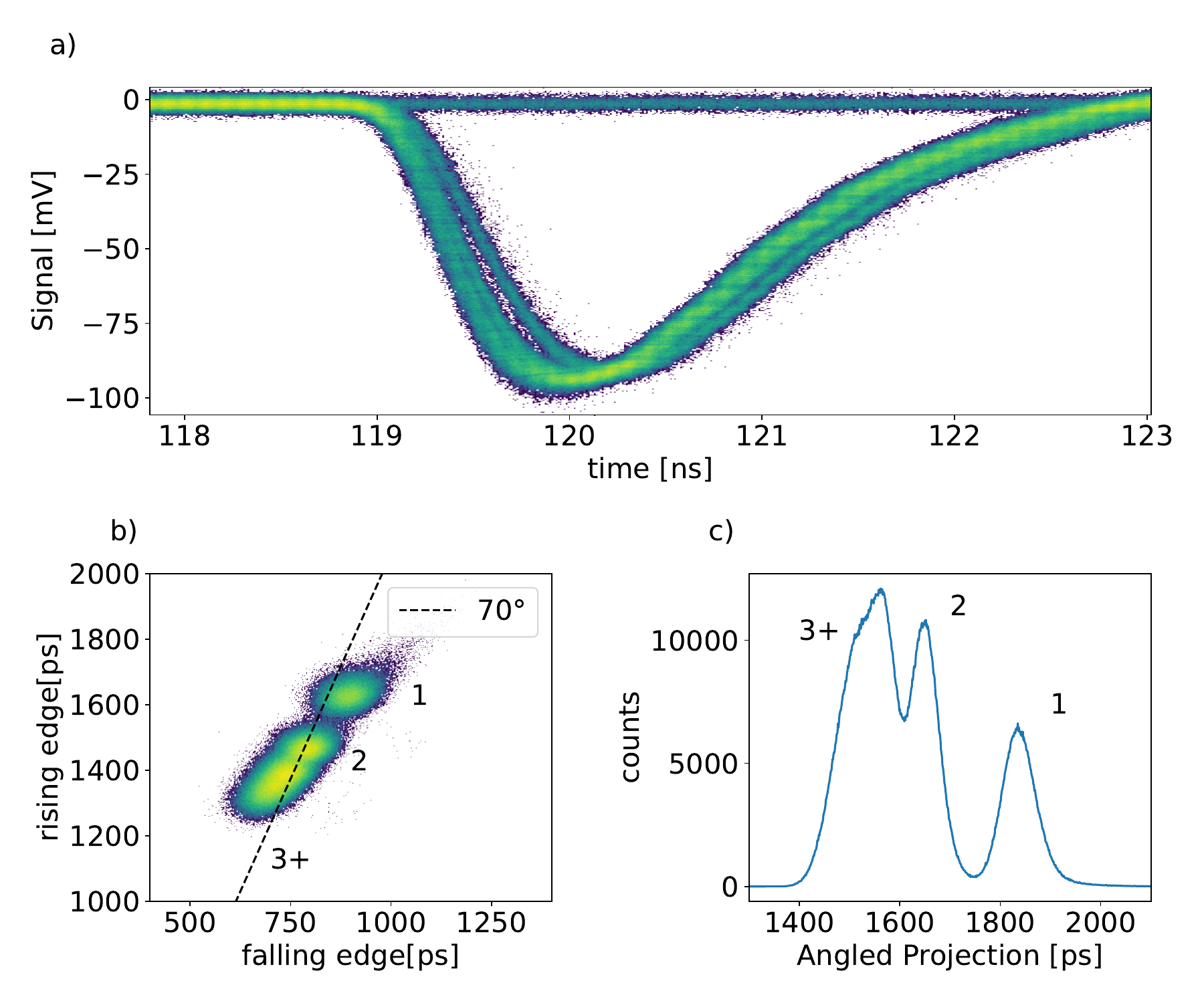}
	\caption{a) A histogram of traces for a single pixel SNSPD illuminated with a coherent state with mean photon number of $\sim$2. b) A two dimensional histogram of delay times between the rising or falling edge of the SNSPD signal and the laser synchronization trigger. The numbers indicate the number of simultaneously detected photons. c) An angled projection of b) along the indicated 70° axis.}
	\label{fig:pnrhists}
\end{figure}

The measurement is performed at a reduced bias current of \SI{10.5}{\micro A} for the detector, as the normally used bias current caused premature latching of the device. This reduced stability was only observed when the commercial cryogenic amplifier was turned on and is likely caused by excess noise generated by this device.

On the oscilloscope, as shown in Fig.~\ref{fig:pnrhists} a), a clear splitting in the falling edges can be seen. The shallower slopes are here indicative of single-photon events, while the faster rising edges were caused by two or more photons hitting the detector at the same time. Another observable difference is a slower falling edge for the single-photon events. It is also possible to see a reduction in peak height for single-photon events. This is likely caused by the second cryogenic amplifier which has a lower bandwidth limit of \SI{200}{MHz} and effectively acts as a high-pass filter. This attenuates the slower and thus (in Fourier-space) lower frequency signal for single-photon detection events. The inversion of the signal is simply due to the amplifier architecture of the first stage amplifier. The features observed here match with other studies of which features indicate PNR~\cite{Schapeler2024}.

\subsection{Application: SNSPD readout with a cryogenic laser}\label{ch:laserreadout}

A second application of the first stage amplifier is its ability to drive other electro-optic components from an SNSPD detection signal. For this proof-of-concept we use a laser diode. This diode allows for the transduction of an electrical signal into an optical signal, that can be transmitted with optical fiber instead of coaxial lines, which reduces heatload from cabling due to the low thermal conductivity of fiber.  More information on the laser readout technique can be found in Ref.~\cite{Thiele2024}.

\begin{figure}
	\centering
	\includegraphics[width=0.7\linewidth]{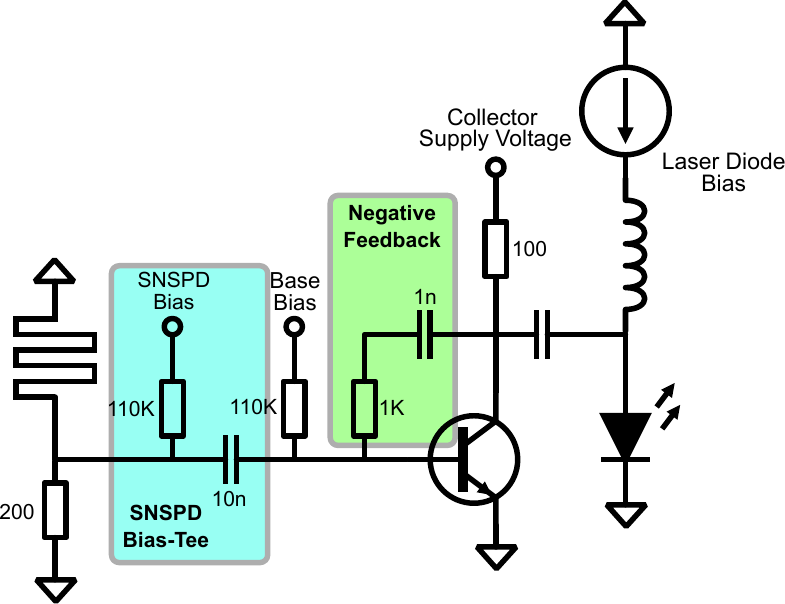}
	\caption{The electrical circuit used to perform optical readout of a SNSPD detection event. The detection signal from the SNSPD is amplified using the low noise amplifier shown in ch. \ref{sec:fsamp} and transmitted to a cryogenic laser diode.}
	\label{fig:laserreadout}
\end{figure}

The electrical circuit for the laser diode based SNSPD readout is shown in Fig.~\ref{fig:laserreadout}. We use the same transistor configuration as in the first stage amplifier (see \ref{sec:fsamp}), but with the output capacitor replaced with a bias tee, to allow for current biasing of the laser diode. This is necessary to increase the modulation speed of the diode, due to faster carrier recombination in the pn-junction during active lasing operation. Biasing of the diode also changes the input impedance, due to the highly nonlinear slope of the diode I-V curve. This can be used to match the input impedance of the laser diode to the output impedance of the amplifier for low reflections.

We also add a \qty{200}{\Omega} resistor near the SNSPD. This acts as an attenuator and prevents early latching of the SNSPD due to additional noise introduced by the laser diode. Despite this the SNSPD still latches at lower bias currents in comparison to conventional operation and also exhibits high dark counts due to stray light from the laser diode.

\begin{figure}
	\centering
	\includegraphics[width=0.8\linewidth]{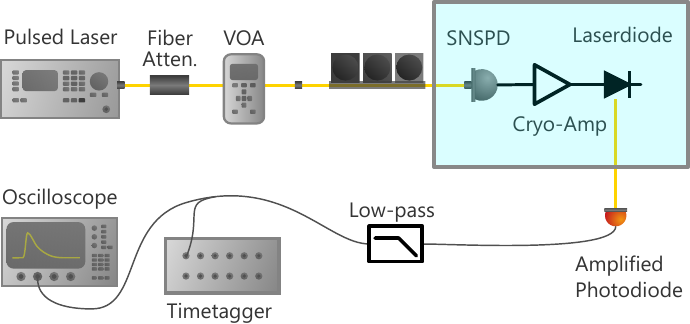}
	\caption{The setup for optical readout of an SNSPD. The SNSPD is triggered using an attenuated laser pulse in the single photon regime. The electrical detection signal is transduced to an optical signal and transmitted to a room temperature photodiode. Based on the signal at the photodiode the SNSPD click signal can be reconstructed.}
	\label{fig:ldsetupsingle}
\end{figure}

\begin{figure}
	\centering
	\includegraphics[width=0.9\linewidth]{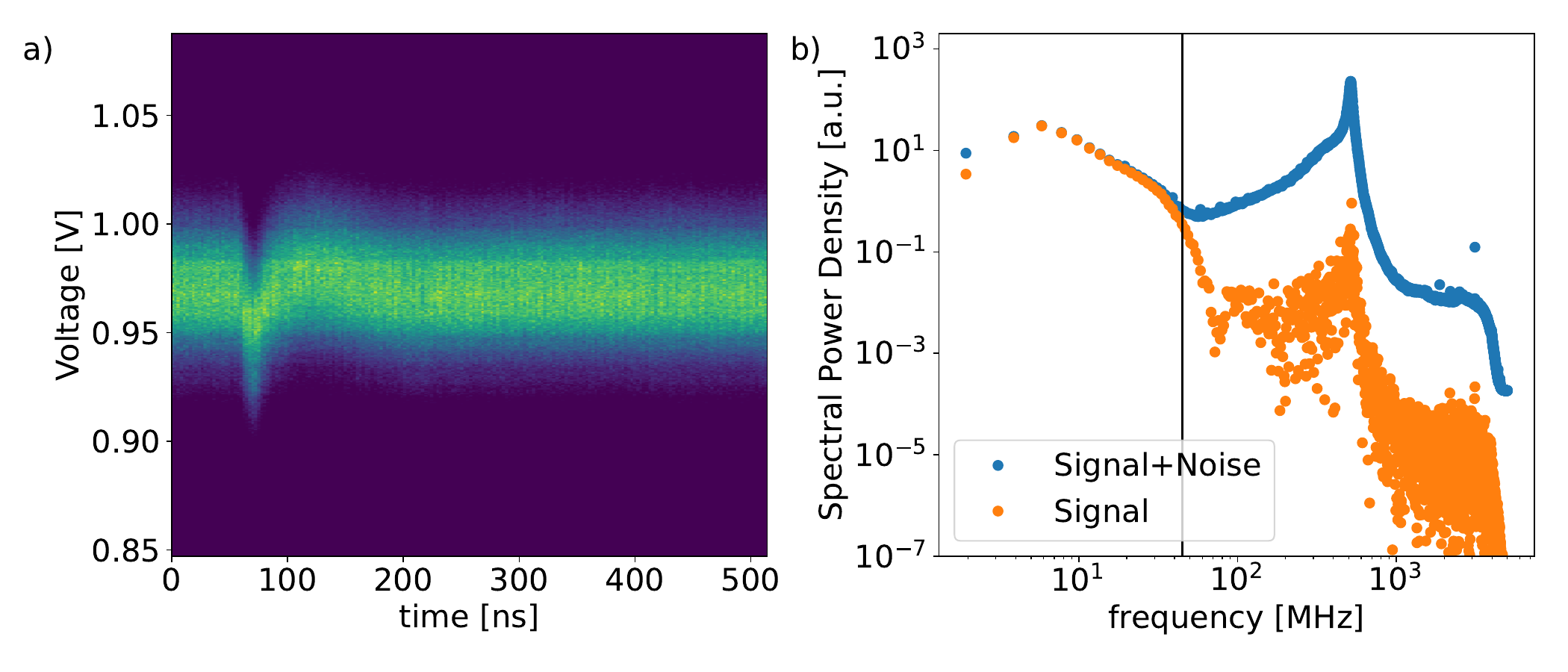}
	\caption{a) The SNSPD signal read out using a cryogenic amplifier and a laser diode. b) A fast Fourier transform of the observed detection signal once without filtering (blue) and once selecting only the periodic elements of the signal (orange), which filter out random noise. The black line indicates the cutoff frequency of the filter used for readout.}
	\label{fig:noisydetetor}
\end{figure}

The setup for measuring the readout signal from the cryogenic laser diode can be seen in Fig.~\ref{fig:ldsetupsingle}. The SNSPD is triggered by a pulsed laser source attenuated to the single-photon regime. The resulting click signal is amplified and converted to an optical signal in the \SI{4}{K} stage of the cryostat and transferred to room temperature by single mode fiber. The resulting optical signal is in the range of \si{\micro W}s and thus needs an amplified photodiode (Finisar FWLF-1519-7D59) for readout. Due to strong high frequency noise, especially around \SI{500}{MHz}, on the optical signal, as shown in Fig.~\ref{fig:noisydetetor}, an additional lowpass filter (Mini circuits VLF-45+ (45 MHz)) was introduced into the readout. This reduces the slope of the rising edge but also reduces noise and therefore makes the signal more usable for edge detection. The source of the excess noise seems to stem primarily from the laser diode, as the noise is drastically decreased when the laser diode is supplied with no current. We presume this might be caused by spectral hole burning at \SI{4}{K} but we have not confirmed this.

The detection signal in Fig.~\ref{fig:noisydetetor}~a) shows an overswing on the falling edge, which indicates a bandwidth mismatch between the SNSPD signal and the readout electronics. This is likely caused by the addition of the laser diode as previous investigations of the amplifier indicate a sufficient bandwidth (see \ref{sec:fsamp}). The signal integrity is however still sufficient to trigger a time-to-digital converter with a degraded jitter of about \SI{1400}{ps} compared to a jitter of \SI{500}{ps} for conventional and \SI{70}{ps} for cryogenic amplification and readout.

\section{Digital Photon-Number Readout} \label{sec:2}

The analog signals generated by SNSPDs can have a high information densities, where photon number and timing information can be encoded in a single waveform. This type of information encoding is, however, incompatible with digital logic gates and modulators, which are key parts of a feed-forward operation. We therefore need to extract information from the analog signal and convert it to a digital signal. %In our case the information we want to extract is the photon number information from the amplitude multiplexed SNSPD-array. 
Here we utilize the previously developed amplifiers to generate a large signal for easier  analysis. The amplified signal is then analyzed with a trigger circuit and additional logic. This is used to discriminate individual peak heights corresponding to reading out the analog signal from an amplitude-multiplexed array of SNSPDs, whose outputs are wired in parallel.

\subsection{Trigger Circuit}
A typical way to perform the transition from analog to digital signals is with a Schmitt-trigger. This circuit switches a digital signal at its output conditional on the input crossing a voltage threshold. One common way to build a Schmitt-trigger is to use an operational amplifier with a positive feedback loop. This is, however, not viable in our cryostat as we have not yet been able to find a commercially available operational amplifier that functions at \SI{4}{K}. This is likely because of the delicate analog design necessary for good operational amplifiers and the frequent use of Si bipolar transistors, that perform poorly at \SI{4}{K}~\cite{Gui2020}. 
We thus need to adapt devices designed for room temperature.

\begin{figure}
	\centering
	\includegraphics[width=0.85\linewidth]{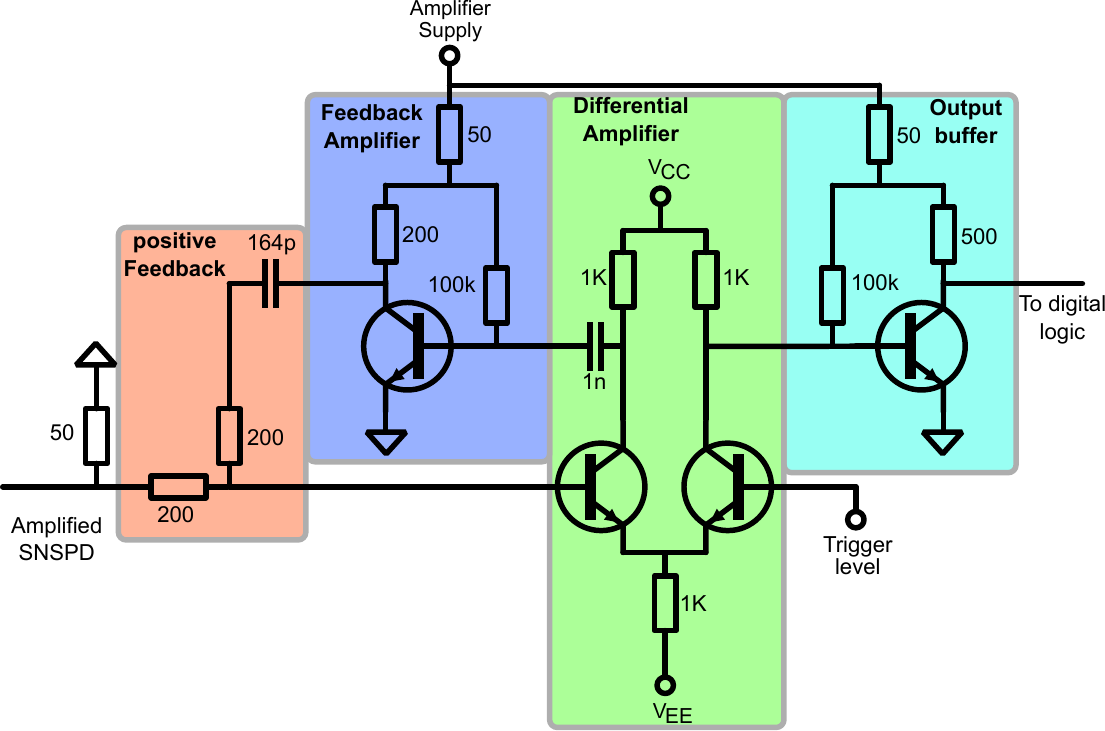}
	\caption{A Schmitt-trigger based on a differential amplifier with an positive feedback loop. The output is buffered with a high impedance output amplifier, to drive subsequent high impedance logic gates.}
	\label{fig:diffampschmitttrigger}
\end{figure}

In order to build a custom Schmitt trigger, we require a circuit that can detect the voltage difference between an SNSPD signal and a triggering threshold, as well as a feedback loop to create a large output signal once this difference becomes positive. The final design for this can be seen in Fig.~\ref{fig:diffampschmitttrigger}. For the differentiation circuit we use a differential amplifier (green), which is made up of two bipolar transistors which share a common current supply at the emitter. A change in voltage at the base of either transistor will increase the current in this transistor and thus decrease the current in the other transistor due to the shared current supply. The amplifier therefore outputs the difference between the voltages at the transistor bases. This can be used to trigger on certain voltage levels by biasing one input at a variable voltage (here labeled Trigger Level). The amplifier will then only create a positive output voltage once this threshold voltage is crossed by the signal coming from the SNSPD.

The signal generated by this type of differentiating circuit is however not able to drive larger currents. We thus employ two additional low noise amplifier circuits (blue) to buffer the outputs and thus increase both the voltage and current generated by the trigger circuit. The stabilisation of the output amplifiers is not done with a resistor-capacitor loop, as used in previous designs (see Fig.~\ref{sec:fsamp}), but instead with a common resistor for both base and collector bias. This acts as a negative feedback loop, which is required to stabilize the amplifier. This configuration was chosen, as in this circuit the collector bias voltage needs to be bigger than the base-bias voltage, due to the voltage requirements of the subsequent logic. This enables this simpler method of introducing negative feedback, which is also more broadband than the capacitor-resistor loop due to the omission of the capacitor.

In order for this circuit to function as a Schmitt-trigger, we require a method to introduce positive feedback between the output and the input of of the differential amplifier.
In this circuit the positive feedback is introduced with a resistor voltage divider (red), to a achieve a 1:1 mixing of the amplified SNSPD signal and the positive feedback at the input. An additional capacitor is introduced into this feedback line. This capacitor charges after a trigger event and thus gradually reduces the amount of positive feedback voltage. It thereby limits the time for which the positive feedback can be active. This decouples the pulse length of the SNSPD from the digital output pulse and thus allows free choice over the length of the subsequent digital signal. 

\subsection{Discrimination Circuit} 
The photon number information of a multiplexed detector is encoded in the amplitude of the generated voltage signal.
Choosing a photon number from this signal can be achieved by looking for peak amplitudes in a given voltage range. To do so we use two trigger circuits with separate triggering levels, where one trigger acts as the lower boundary and the second trigger as the upper boundary. Selecting events between these boundaries thus enables the selection of a desired detected photon number.

\begin{figure}
	\centering
	\includegraphics[width=0.5\linewidth]{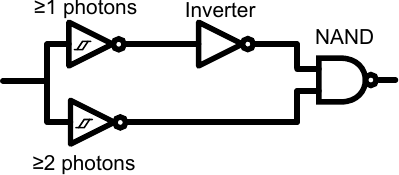}
	\caption{The logic circuit used to select individual photon numbers, here shown in a configuration selecting single photon detection events. The upper trigger acts to detect when the minimum photon number is reached, while the lower trigger blocks the output, when a too high photon number has been detected.}
	\label{fig:logicsheme}
\end{figure}

The logic gates needed to implement this selection can be seen in Fig.~\ref{fig:logicsheme}. We use single photon selection as an example, but the same logic can be applied to select two or three photon events. In order to prevent spurious switching on the rising and falling edges of multi photon events, the signal from the lower boundary Schmitt-trigger is delayed by an inverter and the upper boundary logic pulse is extended by using a larger capacitor in the positive feedback loop. This prevents the lower boundary trigger to activate before the upper boundary has been able to disable the output of the NAND gate.

\begin{figure}
	\centering
	\includegraphics[width=1\linewidth]{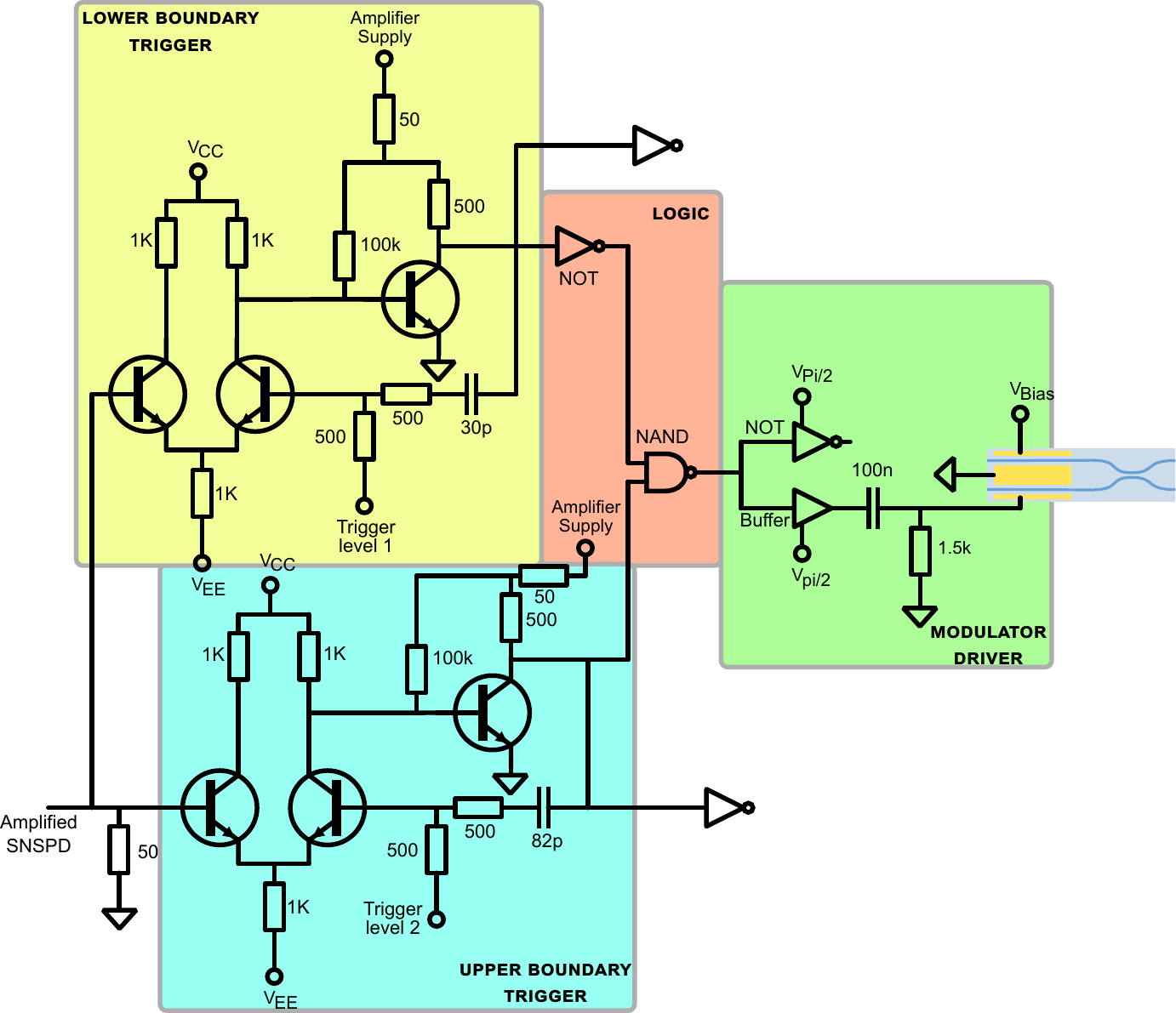}
	\caption{The schematic for the circuit used to demultiplex the amplitude encoded photon number information and generate switching signals based on detecting a given photon number. Additional capacitors for voltage stabilization are not shown. The logic state of the circuit can be probed with the three unconnected NOT gates.}
	\label{fig:fullcircuit}
\end{figure}

The two threshold detectors in this circuit are build identically except for the feedback capacitor, which is larger for the multi photon detector. A capacitor ratio of \SI{30}{pF} to \SI{82}{pF} was chosen; reducing the size of both capacitors can reduce the reset time of the circuit but also reduces the length of the logic signal and therefore the switching window at the modulator. The threshold for both triggers can be set independently with the voltage labeled "Trigger-Level". A full schematic of the resulting electronic circuit can be seen in Fig.~\ref{fig:fullcircuit}

As the last step of the circuit, we implement a driver for a modulator. This would allow the output of discrimination circuit to be used to perform an optical feed-forward operation based. In Thiele et al.,~\cite{Thiele2024b}, we use an electro-optic modulator in based on the titanium in-diffused lithium niobate platform to perform feed-forward based on photon number measurements. Characterization of the modulator has indicated a switching voltage of around \SI{3.8}{V} at \qty{4}{K}.
We can therefore utilize a conventional CMOS inverter and buffer with high current driving capabilities to switch the modulator. This logic gate has been separately tested and found to not meaningfully change at \SI{4}{K} and still support outputs up to \SI{3.6}{V}.

\subsection{Circuit testing}
\subsubsection{Monitored Test}
\begin{figure}
	\centering
	\includegraphics[width=0.7\linewidth]{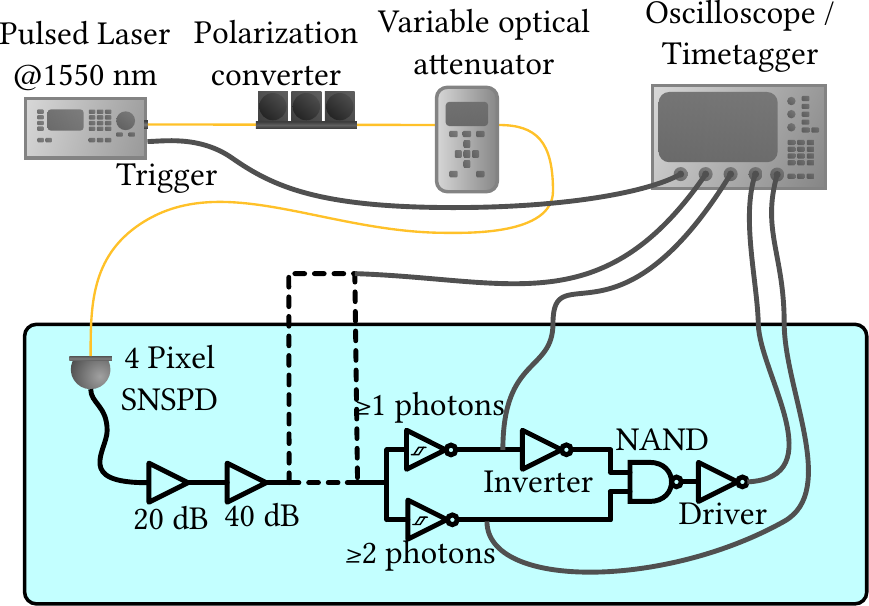}
	\caption{The setup used to characterize the switching behavior of the full cryogenic circuit. The SNSPD is triggered with a pulse laser and subsequently amplified. This amplified signal can be monitored using a cable at room temperature. The SNSPD signal is then sent to the discrimination circuit to generate a photon number selective modulation signal. The cable connection with a dashed line can be made inside or outside the fridge for monitoring the output of the amplifiers.}
	\label{fig:fullcharacterisation}
\end{figure}

We first test the compatibility of the amplified SNSPD with the pulse height discriminator and to check that the modulator driver acts on the correct voltage heights. The full circuit can be operated with an outside connection for monitoring the amplified SNSPD signal. 
The setup used for this can be seen in Fig.~\ref{fig:fullcharacterisation}.

\begin{figure}
	\centering
	\includegraphics[width=0.8\linewidth]{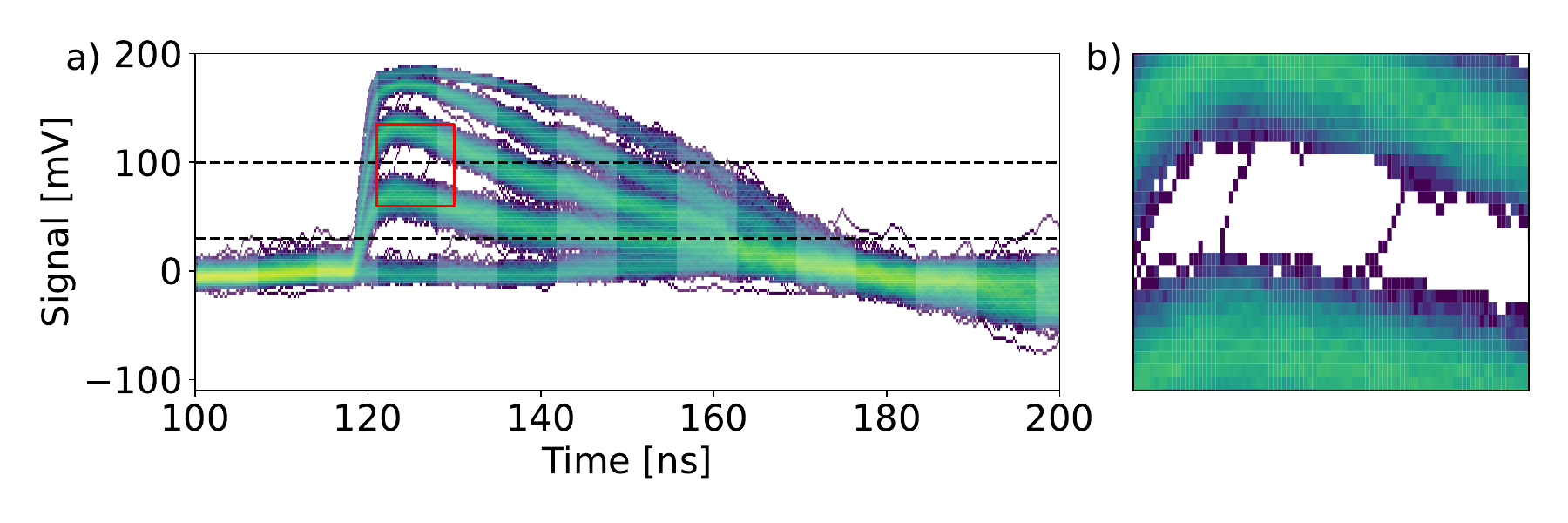}
	\caption{a) A persistence trace of the amplified output of the four pixel SNSPD array, amplified at 4K with the two amplifiers shown previously. The height of the voltage signal indicates the number of detected photons. The dashed lines indicate a possible switching window to select single photon detection events. b) A magnification of the red box in a) showing events where a single photon detection switches over to being a two photon detection.}
	\label{fig:4pixamp}
\end{figure}

We use three amplifier stages based on the design shown in Sec.~\ref{sec:fsamp} %shown in Fig.~\ref{sec:1} 
to amplify the SNSPD click signal to roughly \SI{100}{mV} as can be seen in Fig.~\ref{fig:4pixamp}, which is sufficient for the trigger circuits to switch on. Due to the inverting nature of the amplifiers the SNSPD has to be operated with a negative bias current. This inverts the output pulse of the SNSPD but should not interfere with it otherwise.

One occasional but undesirable observation is the possibility for the detector signal to indicate a single photon detection event in the first few nanoseconds after the laser pulse only for the signal to then rise to a two photon detection event. Two such events can be seen in Fig.~\ref{fig:4pixamp} b). This likely indicates false counts on the detector due to reflections at the first amplifier. This would increase the crosstalk of the detector and is thus undesirable. the probability of observing this effect is however less than 1\%, which is small compared to the normal crosstalk probability of the detector at 13\%~\cite{MATimon}. Further improvements to the reflection coefficient of the first amplifier could be made to reduce this effect.

\subsubsection{Fully Cryogenic Test}
With all individual components tested for functionality and compatibility with each other, the full cryogenic circuit can be assembled.
The full setup used for this test can be seen in Fig.~\ref{fig:fullcharacterisation}, with the SNSPD signal now no longer connected to the outside. 

\begin{figure}[ht]
	\centering
	\includegraphics[width=0.8\linewidth]{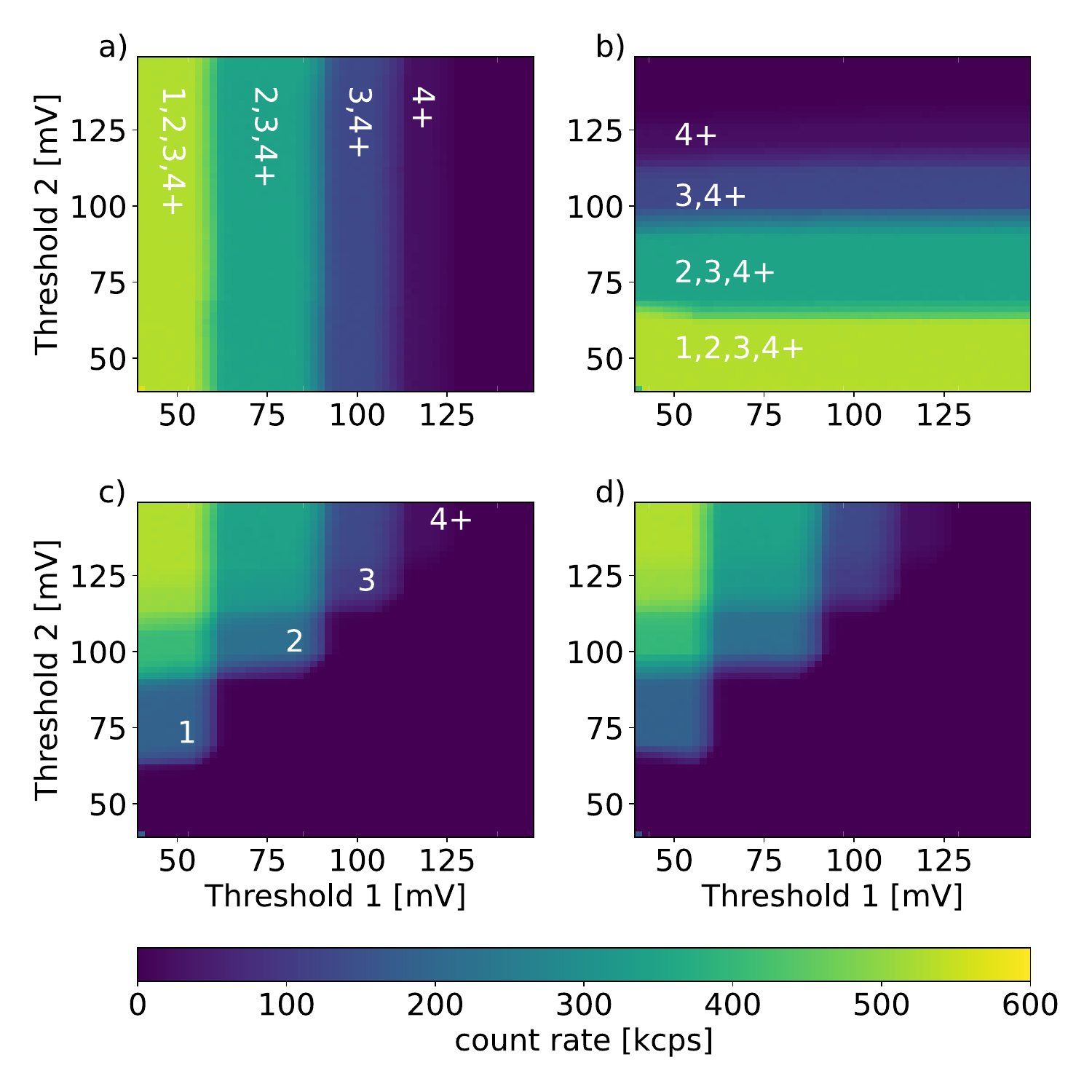}
	\caption{The characterization of the full %feed-forward 
    circuit with pulsed coherent state illumination of the detector. The count rate of the monitoring port for a) the lower boundary trigger (controlled by Threshold 1) and b) the upper boundary trigger (controlled by Threshold 2). c) the count rate at the modulator driver and d) the expected count rate based on the difference of a) and b). The modulator driver shows switching on individual photon numbers as indicated by the white insets.}
	\label{fig:distinctiontest}
\end{figure}

Due to the amplitude multiplexing of the detector it can be expected that a higher triggering threshold will inhibit switching on lower photon numbers. This behavior can be observed in Fig.~\ref{fig:distinctiontest} a) and b). Increasing the threshold level of one trigger circuit reduces the observed count rate, as lower photon numbers detection events are discarded. Changes in the threshold levels only affect the corresponding trigger circuit, while changes to the other triggers threshold do not affect it. This results in four plateau regions for both triggers where the count rates are constant, which indicated good separation between the four amplitude heights coming out of the amplifier chain. These single trigger circuits alone are however not selective on a specific photon number as seen in the inset.

In order to observe switching on individual photon numbers the output of the modulator driver needs to be considered, as shown in Fig.~\ref{fig:distinctiontest} c). Here several plateau regions can be observed, which correspond to different ensembles of photon numbers that the circuit triggers on. The most relevant regions here are indicated with white numbers and represent switching on individual photon numbers. The proper operation of this can be cross-checked by considering the count rate difference between the two individual trigger circuits. The resulting expected count rate can be seen in Fig.~\ref{fig:distinctiontest} d) and matches nicely with Fig.~\ref{fig:distinctiontest} c), confirming the correct functioning of the intermediate logic.

For this circuit, a final estimation of the total heat load and delay can be made. The total heatload based on the power supplied to the circuits is roughly \SI{23}{mW}.
Some of this dissipation is dependent on the photon statistics of the input state and the capacitance of the modulator, as this determines the energy needed to perform a switching operation and the amount of performed switching operations.

The latency of the whole circuit can be estimated based on the latency of all individual components. The length of the wire and traces can be estimated to between 20 cm and 30 cm which leads to a delay of \SI{1.25\pm0.25}{ns} assuming a signal speed of 2/3 the speed of light, which is typical for coaxial cables. For low noise amplifiers the group delay is typically well below a nanosecond and can be ignored in this consideration~\cite{Park2006}. The modulator has been characterized to posses a bandwidth of \SI{230}{MHz} which can be estimated to cause a \SI{10}{\%}-\SI{90}{\%} risetime of \SI{1.5\pm0.5}{ns}. By far the biggest contributor to the delay time is the digital part of the circuit with a delay of \SI{20\pm2}{ns}. Thus a total delay of \SI{23\pm3}{ns} can be estimated. This delay would be faster than room temperature electronics with a \SI{2}{m} coaxial connection out of and back into the cryostat, where the cabling alone would cause roughly \SI{20}{ns} of delay.

\section{Conclusion}
The cryogenic electronic interfacing electronics described in this paper have already enabled a number of experiments, including low heat-load optical readout of an SNSPD~\cite{Thiele2024} and cryogenic feed-forward of a quantum photonic state~\cite{Thiele2024b}. In both cases, it was necessary to build up a library of discrete components and understand how they behave under cryogenic conditions. In particular, this allows us to interface the output of SNSPDs with cryogenic modulators based on the lithium niobate integration platforms.

Up to now, we have been limited to prototype circuitry with discrete components on PCBs. Moving forward, scalability demands implementing these circuits on cryo-compatible integrated chips, which relies on reliable cryogenic CMOS simulation models. Nevertheless, we will continue to expand our library of components to implement proof-of-concept cryogenic circuitry which will enable the next generation of photonic quantum technologies.

\acknowledgments   
 Supported by the Bundesministerium für Bildung und Forschung (BMBF) under the QPIC-1 project (Grant No. 13N15856). Funded by the European Union (ERC, QuESADILLA, 101042399). Views and opinions expressed are however those of the author(s) only and do not necessarily reflect those of the European Union or the European Research Council Executive Agency. Neither the European Union nor the granting authority can be held responsible for them.

\bibliography{library} 
\bibliographystyle{spiebib} 

\end{document}